\documentclass[pre,showpacs,amssymb]{revtex4}
\usepackage{amsmath,graphicx}
\usepackage{epsfig,subfigure}
\usepackage{bm}

\begin{document}

\title{Driven translocation of a polymer: role of  pore friction and
crowding}

\author{J. L. A. Dubbeldam$^{2}$, V. G. Rostiashvili$^1$, A. and T.A.
Vilgis$^1$}
\affiliation{$^1$ Max Planck Institute for Polymer Research, 10 Ackermannweg,
55128 Mainz, Germany\\
$^2$ Delft University of Technology 2628CD Delft, The Netherlands}

\begin{abstract}
Force-driven translocation of a macromolecule through a nanopore is
investigated systematically by taking into account the monomer-pore friction as
well as the
``crowding'' of monomers on the {\it trans} - side of the membrane  which
counterbalance the driving force acting in the pore. The problem is treated 
{\emph self-consistently}, so that the resulting force in the pore and the dynamics on 
the {\it cis} and {\it trans} sides  mutually influence  each other.  The 
set of
governing differential-algebraic  equations for the  translocation dynamics is
derived and solved numerically. The analysis of this solution shows that the 
crowding of  monomers on the trans side hardly affects the dynamics, but the 
monomer-pore friction can
substantially slow down the translocation process. Moreover, the translocation
exponent $\alpha$ in the translocation time - vs. - chain length scaling
law, $\tau \propto N^{\alpha}$, becomes smaller for relatively small chain 
lengths as the  monomer-pore
friction coefficient increases. This is most noticeable for relatively strong 
forces.  Our findings show that the variety of values for  $\alpha$  
reported in experiments and computer simulations, may be attributed to 
different pore frictions, whereas crowding effects can generally be neglected.
\end{abstract}


\maketitle

\section{Introduction}

Force-driven translocation through a nanopore in a membrane is one of the
fastest growing single-molecule manipulation technique \cite{Muthukumar}. The
theoretical
interpretation of this highly nonequilibrium, transient process is mainly based
on the tensile (Pincus) blob picture and the notion of a propagating front of
tensile force along the
chain backbone \cite{Sakaue_1,Sakaue_2,Sakaue_3,Dubbeldam,Grosberg}. In order
to simplify the analysis it was assumed \cite{Sakaue_1,Sakaue_2,Sakaue_3} that
the moving portion of the chain on the  {\it cis} side of the membrane (moving
domain) could be characterized by an average time-dependent velocity $v(t)$. In
other words, the velocity of monomers is the same for every cross-section of the
moving domain; this approximation can therefore be referred to as an {\it
iso-velocity} model. We have earlier pointed out \cite{Dubbeldam} the fact that 
this approximation, although violating
the local material conservation law (continuity equation), could  still 
be used for the integral (or global) conservation law formulation (see the
Appendix in \cite{Dubbeldam}). This then provides a way for a self-consistent
calculation of the chain velocity $v(t)$ which decreases as the tensile front
 propagates.
Moreover, the resulting scaling relationships for the mean translocation time is
compatible with the corresponding result obtained on the basis of the so-called
{\it iso-flux} model \cite{Grosberg} where the flux of monomers is constrained
to be the same through every cross-section of the moving domain.

In this paper we  suggest a consistent generalization  of the tensile force
propagation model by taking into account the   dynamical
effects on the {\it trans} - side of the membrane where a strong crowding of
monomer{\bf s} can be seen \cite{Dubbeldam,Linna_1,Linna_2,Binder}.  It is apparent
that the osmotic
pressure caused by the crowding (which could be quantified in terms of the de
Gennes
concentration blobs \cite{Sakaue_4,DeGennes} ) leads to a  counterbalance of the
driving
force acting in the pore and could result to a slowing  down of the
translocation process \cite{Sakaue_5}. Recently the role of crowding has been
investigated by means of molecular dynamics (MD) simulations of the so-called
``{\it no trans}'' model where a polymer bead is eliminated from the {\it
trans} side as soon as a new bead arrives there \cite{Linna_3}. It has been
shown that such elimination has a very small impact on the translocation
dynamics. Moreover, the role of the polymer-pore friction has been thoroughly
studied  \cite{Ikonen,Ikonen_1,Ikonen_2} using  MD-simulations as well as
the  Brownian dynamics tension propagation (BDTP) model.  It was demonstrated
 that this friction might be a reason for the nonuniversality of the  mean
translocation time $\langle \tau \rangle$ scaling behavior. More precisely, the 
mean translocation time  $\langle \tau
\rangle$ which is generally assumed to scale with chain length $N$ as 
$N^{\alpha}$ with $\alpha$ the 
translocation exponent, was shown to have a value $\alpha$ that decreases with 
pore friction~\cite{Linna_1,Linna_2,Ikonen,Ikonen_2}. 
This of course translates in a dependence of $\alpha$ on the pore size, since  
a smaller
 pore size correponds to a larger pore friction coefficient
\cite{Ikonen_2,Bhattacharya_1,Edmonds}. 
Based on the BDTP - model the finite chain
effect and its impact on the exponent $\alpha$ has been discussed in full
details by Ikonen {\it et al.}
\cite{Ikonen,Ikonen_1,Ikonen_2}. As a result for the force-driven translocation 
(i.e. for the case that the driving force   $f \gg T/aN^{\nu}$, where 
$T$ is the temperature, $a$ is the effective bond length and   $\nu$ is the 
Flory exponent \cite{DeGennes}) for  chain lengths $N$ which
typically used in the simulation or experiments  the translocation exponent
$\alpha$ satisfies the inequality $\alpha < 1 + \nu$, where the value $1+\nu$ 
corresponds to the value of $\alpha$ in the absence of pore friction. 
We recall that for  
unbiased translocation (i.e. translocation without an external driving force) 
this exponent is much larger than $1+\nu$ namely $\alpha = 2\nu + 1$ 
\cite{Kantor} (see also Appendix \ref{App_1}). 

In this paper we generalize the tensile force propagation
model by  explicitly takes into account the dynamics on the {\it trans} - side
(crowding) as well as the polymer - pore friction which  affect the resulting
force in the pore and leads to a more diverse translocation behavior. In doing 
so we treat the problem {\emph self-consistently}. That is, the {\emph effective resulting force} 
in the pore, $F(t)$, has  an impact on the integral material balance of 
polymer segments. On the other hand, force $F(t)$  is affected by the 
dynamics on {\it cis} and {\it trans} sides in a reciprocal manner as it is 
discussed in Sec. II. Our approach extends the BDTP-model 
\cite{Ikonen,Ikonen_1,Ikonen_2} where the time-dependent friction coefficient 
of the {\it cis} side moving domain was taken from the tension-propagation (TP)
model without crowding \cite{Sakaue_1,Sakaue_2,Sakaue_3} supplemented with the 
pore friction. 

In  Sec. II we derive the governing set of equations for  this 
self-consistent translocation model  based
on the tensile blobs on the {\it cis}-side and concentration blobs on the
{\it trans}-side picture. Depending on  driving force one can discriminate
between the ``trumpet, ``stem-flower'' and ``stem'' scenarios. In Sec. III we
solve the resulting equations numerically and discuss in detail how the
translocation exponent depends on driving force, pore friction and  chain
length. We  conclude with an extended summary of our results in Sec. IV.

\section{Single chain dynamical response}

\subsection{Tensile-blob picture on the {\it cis} - side}

\begin{figure}[ht]
 \includegraphics[scale=0.45]{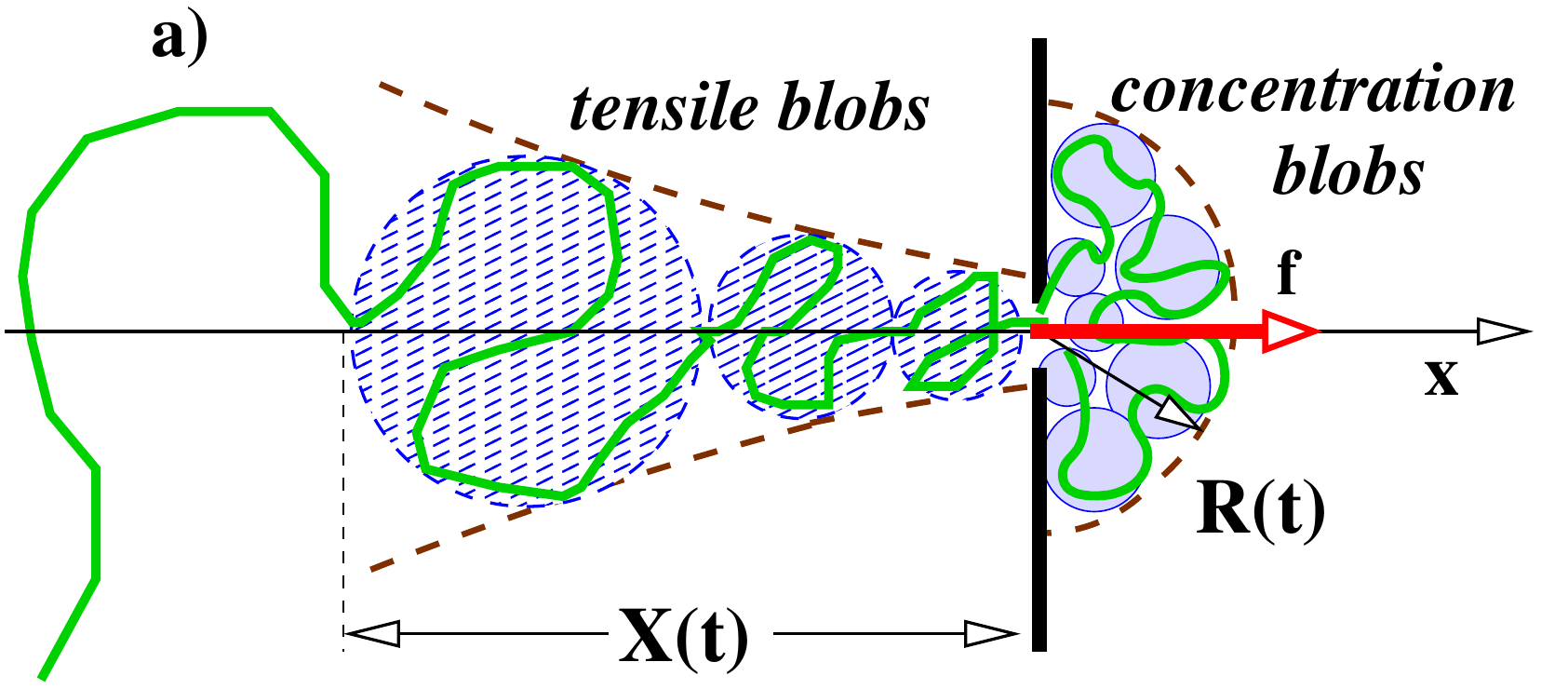}
\hspace{0.8cm}
\includegraphics[scale=0.45]{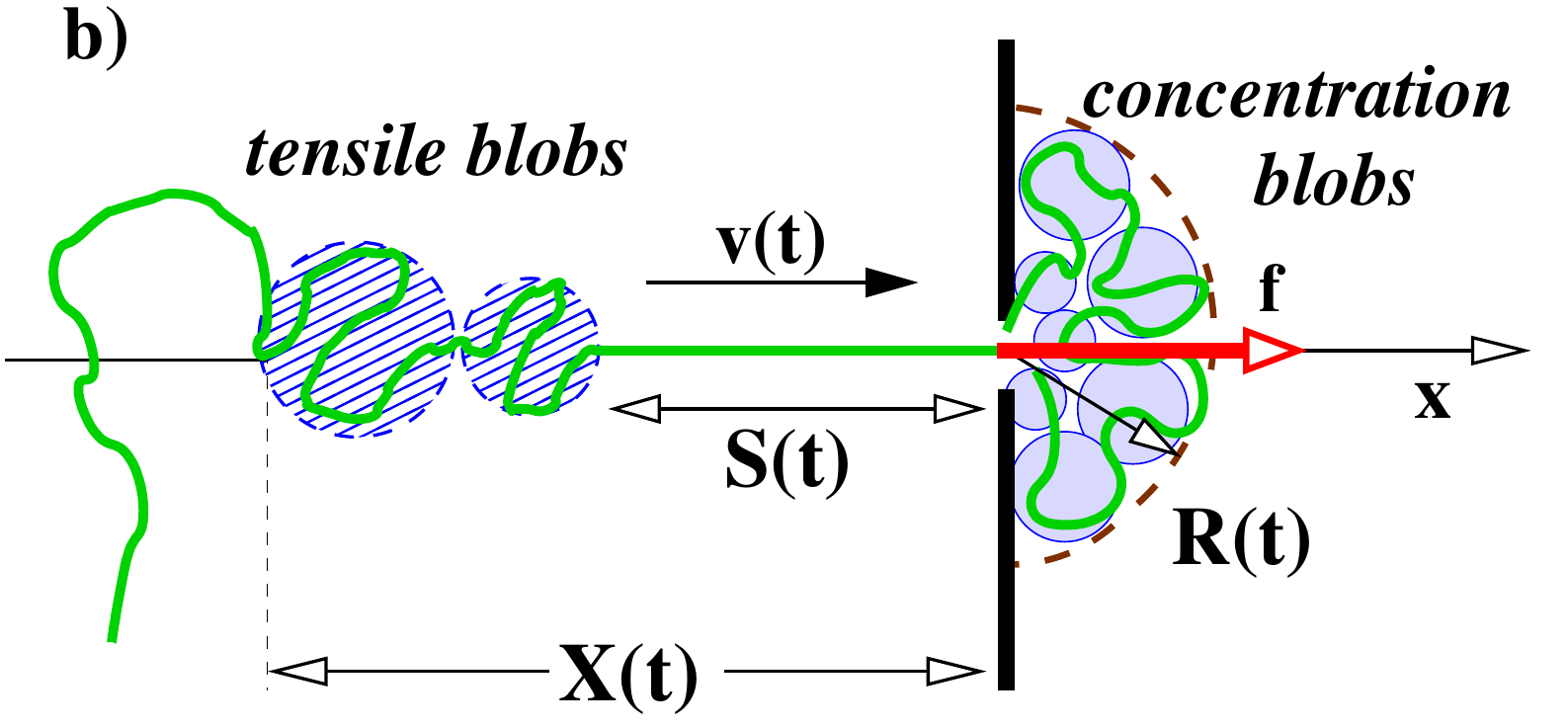}
\vspace{1cm}
\includegraphics[scale=0.45]{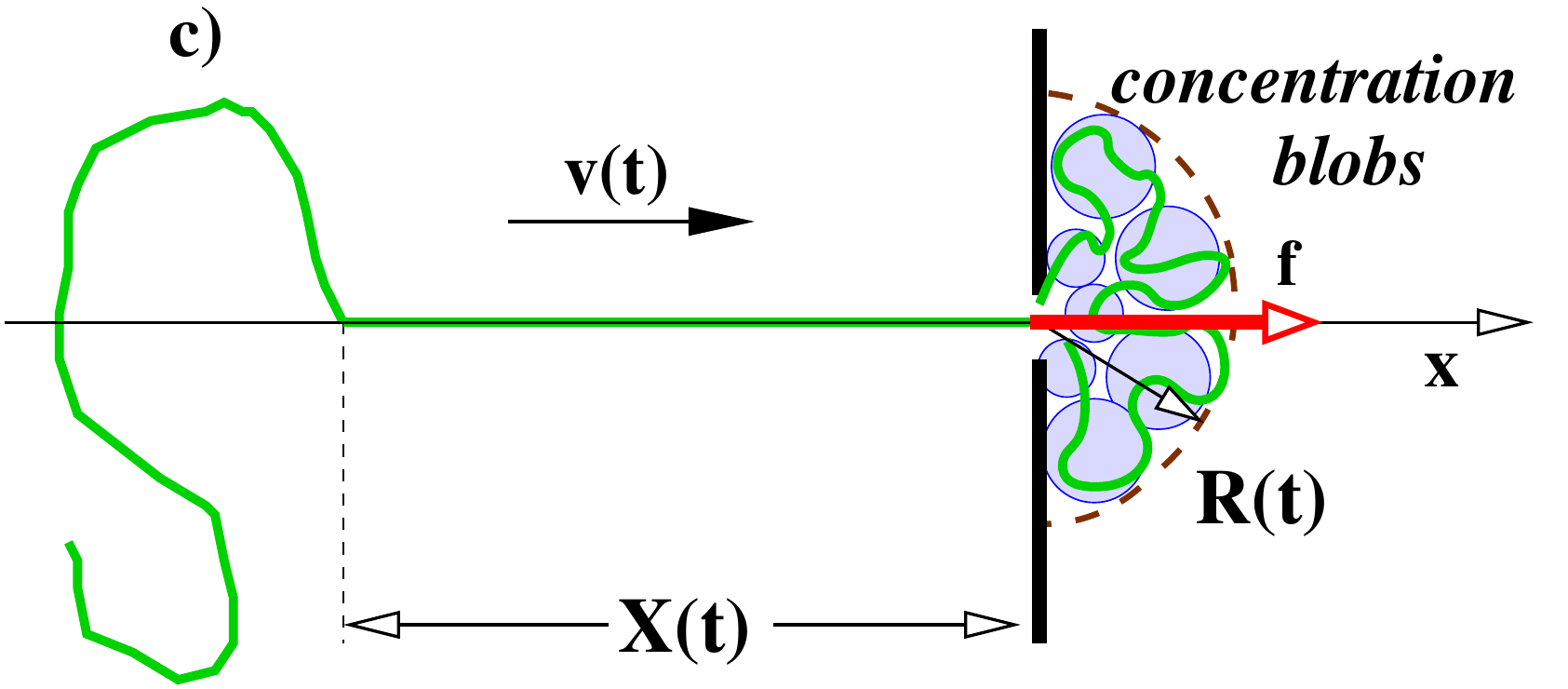}
\caption{Dynamical response of the polymer chain to the driving force $f$ acting
in the pore. Concentration blobs
on the {\it trans} side make up a
``mushroom cap''. Blob sizes  on the {\it cis} - side  are $x$-dependent and
have
the cylindrical symmetry, while the concentration blobs have a hemispherical
symmetry. The front propagation on the {\it cis}- and {\it trans}- sides are
denoted by $X(t)$ and $R(t)$ respectively. (a)  For small forces, $T/a N^{\nu}
\ll f \le T/a$, tensile blobs on
the {\it cis} - side are  shaped in the form of
a ``trumpet'' ;  (b) For the intermediate forces, $T/a < f \ll (T/a)
N^{\nu}$, the part of the chain affected by tension  has a
``stem-flower'' configuration, with the stem length $S(t)$; (c) Finally, for
large forces, $f > (T/a) N^{\nu}$, the
part of chain affected by tension looks like a ``stem''. }
\label{Schematic}
\end{figure}

On the {\it cis} - side of the membrane the moving domain has a cylindrical
symmetry and the tensile (or Pincus) blobs  are shaped in the form of a
``trumpet'' as it is pictured in Fig. \ref{Schematic}. As usual the trumpet
regime takes place for  moderately strong driving forces  $f$ falling within the
range $T/a N^{\nu} \ll f \le T/a$, where $N$ is the chain length, $T$ is the
temperature  and $\nu$
stands for the Flory exponent. In Fig. \ref{Schematic} the distance between the
propagating tension front and the membrane is marked as $X(t)$. 

\subsubsection{Weak forces: $T /a N^{\nu} \ll f \leq T/a$}

The tensile blob size $\xi (x, t)$ , located at  distance $x$ from the
membrane, 
can be expressed in terms of the corresponding tensile force $f(x, t)$ as
follows
\begin{eqnarray}
 \xi (x, t) = \dfrac{T}{f(x, t)}.
\label{Blob}
\end{eqnarray}

On the other hand, the local tensile force $f(x, t)$  is balanced by the
Stokes friction force acting on the segments located between $x$ and $- X(t)$
(recall that the origin of coordinates, $x = 0$, is placed in the pore ), i.e.
\begin{eqnarray}
 f(x, t) = 6\pi \: \zeta_0  \int\limits_{-X(t)}^{x}  v (x', t) \left[ \dfrac{\xi
(x', t)}{a} \right]^{z - 2} \: \dfrac{d x'}{\xi (x', t)}.
\label{Balance}
\end{eqnarray}
In Eq. (\ref{Balance})  $d x'/\xi (x', t)$ counts the number of blobs in the
interval $x', x' + dx'$, whereas $\zeta_0 v (x', t)  [\xi (x', t)/a]^{z - 2}$ is
the local Stokes friction force, with $\zeta_0$ being the friction coefficient
\cite{DeGennes}.  The dynamic exponent is equal either $z = 2 + 1/\nu$  or $z
= 3$ for Rouse or Zimm dynamics respectively. We also keep  explicitly the
coefficient $6 \pi$  bearing  in mind that in both, Rouse or  Zimm,  cases we
have a succession of spherical units, beads or impenetrable blobs,  which are
responsible for Stokes friction. 

In order to simplify the mathematical treatment of the tensile propagation (on
the {\it cis} - side) and the crowding effect on the {\it trans} side as well
as to gain more physical insight into the process we will rely on the
``homogeneous approximation''. In this case the tensile propagation
consideration is based on the ``two phase picture'' \cite{Sakaue_1,Sakaue_2}
where the moving and quiescent domains are separated by a narrow tension front.
Moreover, the moving domain is treated as a uniform block within which all
monomers are moving with the same time-dependent representative velocity
$v(t)$. In this case instead of Eq. (\ref{Balance}) we have
\begin{eqnarray}
 f(x, t) = 6 \pi \: \zeta_0 \: v (t) \int\limits_{-X(t)}^{x}   \left[ \dfrac{\xi
(x', t)}{a} \right]^{z - 2} \: \dfrac{d x'}{\xi (x', t)}.
\label{Balance_Homogen}
\end{eqnarray}

By making use the local relationships $\xi (x, t) = T/f(x, t)$ we have
\begin{eqnarray}
 \dfrac{1}{\xi (x, t)} = \dfrac{6 \pi \:\zeta_0}{T} \: v (t)
\int\limits_{-X(t)}^{x}  
\left[ \dfrac{\xi
(x', t)}{a} \right]^{z - 2} \: \dfrac{d x'}{\xi (x', t)},
\end{eqnarray}
which when recast in differential form reads
\begin{eqnarray}
 \dfrac{d \xi}{d x} = - \dfrac{6 \pi \:\zeta_0 \: v(t)}{T a^{z-2}} \: \xi^{z -
1}.
\label{Blob_Dif_Equation}
\end{eqnarray}
Equation \ref{Blob_Dif_Equation} should be  supplemented with the boundary
condition
\begin{eqnarray}
 \xi (x=0, t) = \dfrac{T}{F(t)},
\end{eqnarray}
where $F(t)$ stands for the resulting force in the pore (see below for more
details). As a result the solution reads
\begin{eqnarray}
 \xi (x, t) = \dfrac{a}{\left\lbrace   \dfrac{6 \pi \:\zeta_0 v(t)}{T} \: x +
\left[\dfrac{a F(t)}{T} \right]^{z-2}\right\rbrace^{1/(z - 2)}}
\label{Blob_Solution}
\end{eqnarray}
The second  condition at the free boundary, i.e. at $x = - X(t)$, claims that 
the tension $f(x = - X(t), t) = T/ \xi (x = -X(t), t) = 0 $. Taking into
account Eq. (\ref{Blob_Solution}) we have $- 6 \pi \:\zeta_0 v(t) X(t)/T + [a
F(t)/T]^{z-2} = 0$ and hence the chain velocity
obeys\begin{eqnarray}
 {\widetilde v}(t) = \dfrac{[{\widetilde F}(t)]^{z-2}}{{\widetilde X}(t)},
\label{Velocity}
\end{eqnarray}
where we introduced the dimensionless quantities: ${\widetilde v}(t) \equiv 6 \pi \:\zeta_0 a v(t)/T$, ${\widetilde F}(t)
\equiv 
a F(t)/T$ and ${\widetilde X}(t) \equiv X(t)/a$. Taking Eq. (\ref{Velocity})
into account Eq. (\ref{Blob_Solution}) could be represented in the form
\begin{eqnarray}
 {\widetilde \xi} ({\widetilde x}, t) = \dfrac{1}{\{{\widetilde v}(t)
[{\widetilde x} + {\widetilde X}(t)]\}^{1/(z-2)}},
\label{Blob_2}
\end{eqnarray}
where the notations ${\widetilde x} \equiv x/a$ and ${\widetilde \xi}  \equiv
\xi/a$ have been used.

Using  the global material balance of monomers we arrive at the following relation
\begin{eqnarray}
 \int\limits_{- X(t)}^{0} \: \left[\dfrac{\xi (x, t)}{a} \right]^{1/\nu}
\dfrac{d x}{\xi (x, t)} + M(t) = N(t),
\label{Material_Balance}
\end{eqnarray}
where the first integral counts the number of monomers in the moving domain
(see Fig. \ref{Schematic}), $M(t)$ denotes the number of translocated monomers
and $N(t)$ stands for the total number of monomers subjected to tension during
the time interval $[0, t]$. At time $t = 0$ all these $N(t)$ monomers were
in equilibrium and occupying a region of size $X(t)$. In other words $N(t)$ 
and $X(t)$ are related by the Flory expression, i.e.
\begin{eqnarray}
 X(t) = a [N(t)]^{\nu}.
\label{Flory}
\end{eqnarray}
Substituting  Eq.~(\ref{Blob_2}) and Eq.~(\ref{Flory}) in
 Eq.~(\ref{Material_Balance}) for the material balance, leads to
\begin{eqnarray}
 \dfrac{{\widetilde X}(t)}{ C_1 \: [{\widetilde F}(t)]^{\frac{1}{\nu} - 1}} +
M(t) = [{\widetilde
X}(t)]^{\frac{1}{\nu}},
\label{Mat_Balance_Final}
\end{eqnarray}
where the numerical coefficient $C_1 = 1 - (1 - \nu)/[\nu (z - 2)]$, i.e. $C_1
= \nu$ for Rouse and $C_1 = 2 - 1/\nu$ for Zimm models.

The flux of monomers at the pore $j_0 (t) = \rho (x = 0, t) v(t)$ (where
$\rho (x, t)  = [\xi (x, t)/a]^{1/\nu}/\xi(x, t))$ is the linear density of
monomers) should be taken equal to $d M(t)/dt$ , i.e. as a result 
\begin{eqnarray}
 \dfrac{d M(t)}{d t} = \left[ \dfrac{\xi (x = 0, t)}{a} \right]^{1/\nu}
\dfrac{v(t)}{\xi (x = 0, t)}\nonumber,
\end{eqnarray}
or in terms of the dimensionless variables
\begin{eqnarray}
\dfrac{d M(t)}{d{ \widetilde t}} = \dfrac{[{\widetilde F}(t)]^{z-1 -
1/\nu}}{ 6\pi \: {\widetilde X}(t)}.
\label{Flux_of_Monomers}
\end{eqnarray}
where we have  used Eq. (\ref{Velocity}) and Eq. (\ref{Blob_2}) and also
introduced the dimensionless time ${ \widetilde t} = t/\tau_0$, with $\tau_0 =
a^2 \zeta_0/T$.

It is worth mentioning  that the foregoing consideration refers to the 
tension propagation. After the characteristic time $\tau_1$ when the tension
has propagated to the last monomer of the chian, i.e. at ${\widetilde X}(\tau_1) = N^{\nu}$ or
$N(\tau_1) = N$, the second, so-called tail retraction, stage sets in. For $t > \tau_1$ the  material balance equation
(\ref{Mat_Balance_Final})  should therefore be replaced by the following relation
\begin{eqnarray}
 \dfrac{{\widetilde X}(t)}{ C_1 \: [{\widetilde F}(t)]^{\frac{1}{\nu} - 1}} +
M(t) = N.
\label{Mat_Balance_Final_1}
\end{eqnarray}

In the case of intermediate  driving forces, i.e. for $ (T/a) < f \ll T/a$, the
foregoing 
equations Eqs.~(\ref{Velocity}), (\ref{Mat_Balance_Final}),
(\ref{Flux_of_Monomers}), (\ref{Mat_Balance_Final_1}), will change form, which we will
discuss next.

\subsubsection{Intermediate forces: $ T/a < f \ll (T/a) N^{\nu}$}

In this case the translocation starts with a ``stem'' formation and the
velocity decreases , so that at the moment $t = \tau^{\sharp}$ the drag force at
the stem-flower junction point becomes $T/a$, i.e. $6\pi \zeta_0
v(\tau^{\sharp}) = T/a$ or in dimensionless notations ${\widetilde v}
(\tau^{\sharp}) = 1$. At $t > \tau^{\sharp}$ the ``stem-flower'' regime sets in
(see Fig. \ref{Schematic}b) with the ``flower'' part following the same as for
the weak force differential equation, Eq. (\ref{Blob_Dif_Equation}). However,
the boundary condition is different and reads $\xi (x = - S(t)) = a$. Thus, the
``flower'' part follows the law 
\begin{eqnarray}
 \xi ({\widetilde x}, t) =  \dfrac{a}{\left\lbrace   1 + {\widetilde v} (t) [ 
{\widetilde x} + {\widetilde S}(t)]\right\rbrace^{1/(z-2)}},
\label{Flower_Law}
\end{eqnarray}
where the dimensionless values ${\widetilde x} = x/a$ and ${\widetilde
S}(t) = S(t)/a$. Again at ${\widetilde x}  = - {\widetilde X}(t) $ the tensile
force is zero, i.e. $f({\widetilde x} = - {\widetilde X}(t)) = T/\xi
({\widetilde x} = - {\widetilde X}(t), t) = 0$ and by making use Eq.
(\ref{Flower_Law}) we have
\begin{eqnarray}
 {\widetilde X}(t) = {\widetilde S}(t) + \dfrac{1}{{\widetilde v} (t)}.
\label{S_vs_X}
\label{Velocity_Law}
\end{eqnarray}

The material balance is in this case given by ({\it cf.} Eq. (\ref{Material_Balance}))
\begin{eqnarray}
  \int\limits_{- {\widetilde X}(t)}^{- {\widetilde S}(t)} \: \left[\dfrac{\xi
({\widetilde x}, t)}{a} \right]^{1/\nu}
\dfrac{ a \: d {\widetilde x}}{\xi ({\widetilde x}, t)} + {\widetilde S}(t) +
M(t) = N(t),
\label{Material_Balance_Stem_Flower}
\end{eqnarray}
which after  using  Eqs. (\ref{Flory}) and (\ref{Flower_Law}) takes the form
\begin{eqnarray}
 \dfrac{1}{C_1  {\widetilde v}(t) } + {\widetilde S}(t) +  M(t) = [{\widetilde
X}(t)]^{1/\nu}.
\label{Material_Balance_Stem_Flower_1}
\end{eqnarray}
To exclude ${\widetilde v}(t)$ and ${\widetilde S}(t)$ we write down the force
balance for the stem, i.e. $6\pi \zeta_0 v(t) {S}(t)/a = F (t) - T/a$
or in terms of dimensionless variables ${\widetilde v}(t) {\widetilde S}(t) =
{\widetilde F}(t) - 1$. Combination of this result with Eq.
(\ref{Velocity_Law}) leads to
\begin{eqnarray}
 {\widetilde v}(t) = \dfrac{{\widetilde F}(t)}{{\widetilde X}(t)},
\label{Velocity_Stem_Flower}
\end{eqnarray}
and
\begin{eqnarray}
 {\widetilde S}(t) = {\widetilde X}(t) - \dfrac{{\widetilde X}(t)}{{\widetilde
F}(t)}.
\label{Stem_Length}
\end{eqnarray}
Thus, the material balance Eq. (\ref{Material_Balance_Stem_Flower_1}) becomes
\begin{eqnarray}
 \left( \dfrac{1}{C_1} - 1\right) \dfrac{{\widetilde X}(t)}{{\widetilde F}(t)}
+ {\widetilde X}(t) + M(t) = \left[ {\widetilde X}(t) \right]^{1/\nu}
\label{Material_Balance_Stem_Flower_Final}
\end{eqnarray}
For the same reason as in the weak force case Eq.~(\ref{Material_Balance_Stem_Flower_Final}) only applies for $t~\leq~\tau_1$.
For $t >\tau_1$  Eq. (\ref{Material_Balance_Stem_Flower_Final}) should be
must be replaced by the expression
\begin{eqnarray}
 \left( \dfrac{1}{C_1} - 1\right) \dfrac{{\widetilde X}(t)}{{\widetilde F}(t)}
+ {\widetilde X}(t) + M(t) = N.
\label{Material_Balance_Stem_Flower_Final_1}
\end{eqnarray}

The flux of monomers through the pore is $d M (t)/d t = v(t)/a$ or in
fully dimensionless variables this reads
\begin{eqnarray}
 \dfrac{d M({\widetilde t})}{d {\widetilde t}} = \dfrac{{\widetilde
F}({\widetilde t})}{6 \pi {\widetilde X}({\widetilde t})},
\label{Flux_of_Monomers_Stem_Flower}
\end{eqnarray}
where we have invoked Eq.~(\ref{Velocity_Stem_Flower}). We next turn to the strongly forced chain.

\subsubsection{Strong forces: $ f >  (T/a) N^{\nu}$}
\label{Intermediate_Force}

In this case the moving domain on the {\it cis}-side is completely stretched
(``stem'') as
shown in Fig. \ref{Schematic}c and the force balance reads
\begin{eqnarray}
 {\widetilde F}(t) = {\widetilde v}(t) {\widetilde X}(t).
\label{Balance_Force}
\end{eqnarray}

The material balance for $t < \tau_1$  is simply given by
\begin{eqnarray}
 {\widetilde X}(t) + M(t) = N(t).
\label{Material_Balance_Stem}
\end{eqnarray}
Taking into account again  that $N(t)^{\nu} = {\widetilde X}(t)$ we have
\begin{eqnarray}
 {\widetilde X}(t) + M(t) = [{\widetilde X}(t)]^{1/\nu}.
\label{Material_Balance_Stem_1}
\end{eqnarray}

For $t > \tau_1$ the material balance takes the form
\begin{eqnarray}
 {\widetilde X}(t) + M(t) = N.
\label{Material_Balance_Stem_2}
\end{eqnarray}
where $N$ is the chain length.

Equation for $M(t)$ has the following form (in dimensionless variables) $d
M(t)/d {\widetilde t} = {\widetilde v} (t)/6\pi$. 
Taking into account the force balance equation, Eq.~(\ref{Balance_Force}) we 
arrive at
\begin{eqnarray}
 \dfrac{d M(t)}{d {\widetilde t}} = \dfrac{{\widetilde F} (t)}{6 \pi {\widetilde
X}(t)},
\label{Flux_of_Monomers_Stem}
\end{eqnarray}
which is exactly equivalent to the corresponding Eq.
(\ref{Flux_of_Monomers_Stem_Flower}) for the ``stem-flower'' case.
It is also interesting that this equation exactly corresponds to 
Eq.~(\ref{Flux_of_Monomers}) taken for the Rouse model, i.e. at $z-2 = 1/\nu$.

The two equations, Eq.~(\ref{Mat_Balance_Final}) (or the corresponding 
Eq.~(\ref{Mat_Balance_Final_1})) and Eq. (\ref{Flux_of_Monomers}),
for two unknowns, ${\widetilde X}(t)$ and $M(t)$, are  still not closed, because
the resulting force ${\widetilde F}(t)$ acting in the pore is not simply a given
function of time. This force includes the driving force $f$  which is balanced
by the pore friction as well as  the osmotic pressure on the {\it trans}- side 
(crowding effect).
In order to quantify the last one we should investigate the blob dynamics on
the {\it trans}-side in more detail, to which we turn in the next subsection.

\subsection{Concentration-blob picture on the {\it trans}-side}

In the ``homogeneous approximation'' used before, the monomer density in the
hemisphere of size $R(t)$ is uniform and mass density $ \phi (t) = a^3 M(t)/[R(t)]^3$. 
The concentration blob size $\xi(t)$ is now given by
\begin{align}
\xi (t) &= a [\phi (t) ]^{\nu/(1 - 3\nu)} = a
[ {\widetilde R}(t)^3/M(t)]^{\nu/(3\nu - 1)},
\label{joh1}
\end{align}
 where the dimensionless
${\widetilde R}(t) \equiv R(t)/a$ was introduced. This approximation in the context of polymer
decompression dynamics has been previously discussed by Sakaue {\it et al.}
\cite{Sakaue_4}.

We next derive the differential equation for ${\widetilde R}(t)$. The
confinement free energy $\Delta {\cal F}$ can be written as a number of concentration blobs, 
$R(t)^3/\xi (t)^3$, times the temperature $T$ \cite{Sakaue_4,DeGennes}. If we next use Eq.~(\ref{joh1}) we find
\begin{eqnarray}
 \Delta {\cal F} = T \dfrac{R(t)^3}{\xi (t)^3} = T \: C_2 \: \left[ 
\dfrac{M(t)^{\nu}}{{\widetilde R}(t)}\right]^{3/(3\nu - 1)},
\label{Delta_Free_Energy}
\end{eqnarray}
where $C_2$ is a  constant of order unity.
The equation of motion for ${\widetilde R}(t)$ can be obtained by equating
the friction (or drag) force $f_{\rm fr}$ to the thermodynamic force $f_{\rm
th} = - \partial \Delta {\cal F}/\partial R$. In the Rouse model the chain is
fully free-draining and all beads experience the same friction, i.e. $f_{\rm fr} =
6 \pi \zeta_0 M(t) d R(t)/d t$, where $\zeta_0$ is a monomer  friction
coefficient.
In the Zimm model the friction force is defined by the geometric dimension of
the {\it trans}-domain  times the velocity, i.e. $f_{\rm fr} = 6 \pi \eta_0 R(t) d
R(t)/d
t$, where $\eta_0 \sim  \zeta_0/a$ is the solvent viscosity. The thermodynamic
force $f_{\rm th}$ is given by
\begin{eqnarray}
 f_{\rm th} = 
- \dfrac{\partial}{\partial R} \: \Delta
{\cal F} \simeq  T \left(  \dfrac{3 C_2}{3 \nu - 1}\right) \dfrac{  [M]^{3\nu/
(3\nu - 1)}}{a [{\widetilde
R}]^{(3\nu + 2)/(3\nu-1)}} 
\label{Th}
\end{eqnarray}
Balancing the friction and thermodynamic forces, $f_{\rm fr} = f_{\rm th}$, we find the 
governing equation for $\widetilde{R}$ in the Rouse and Zimm case as
\begin{eqnarray}
  \dfrac{d {\widetilde R}}{d {\widetilde t}} = \begin{cases}
                  \dfrac{C_2 [M]^{1/(3\nu - 1)}}{2 \pi (3\nu - 1){\widetilde
R}^{(3\nu+2)/(3\nu-1)}}
, &\mbox{Rouse} \\
\\
                 \dfrac{ C_2 [M]^{3\nu/(3\nu - 1)}}{2 \pi (3\nu - 1){\widetilde
R}^{(6\nu+1)/(3\nu-1)}} ,&\mbox{Zimm} 
 \label{R}                \end{cases}
\end{eqnarray}

Equation~(\ref{R}) is usually referred to as the Onsager equation \cite{Puri}.
Finally, we discuss the resulting force $F(t)$ acting in the pore.

\subsection{Resulting force $F(t)$ in the pore}

The driving force $f$ push the monomers in the {\it trans}-domain which has a 
hemispherical form of size $R(t)$. In the ``homogeneous approximation'' this 
process could be seen as the work done against the osmotic pressure within the 
hemisphere, i.e. the monomer in the pore which is about to translocate  could 
be thought of as a small piston. In other words, the driving force is 
counterbalanced by the osmotic pressure times the pore cross-area.  
Thus, the resulting force $F(t)$  in the pore  is made up of following  
components: the external {\it driving force} $f$ which
is mitigated by the {\it counterbalance force} $f_{\rm count} (t)$, caused by
the osmotic pressure in the compressed {\it trans}-domain (crowding effect), as 
well as by the the
friction force $f_{\rm pore}(t)$ in the pore, i.e.
\begin{eqnarray}
 F(t) = f - f_{\rm count}(t) - f_{\rm pore}(t).
\label{F}
\end{eqnarray}
 The force $f_{\rm count} (t)$ is defined as the osmotic pressure times the
cross-sectional  area of the pore, i.e.
\begin{eqnarray}
 f_{\rm count} (t) = \underbrace{\left(- \dfrac{\partial \Delta {\cal
F}}{\partial V} \right)}_{\rm osmotic
\: \: pressure} \times \underbrace{\dfrac{\pi}{4} a^2}_{\rm area} = \dfrac{T}{a}
\left[ \dfrac{3 C_2}{8(3\nu - 1)} \right]\left(
\dfrac{M}{{\widetilde R}^3}\right)^{3\nu/(3\nu - 1)},\nonumber
\end{eqnarray}
where we took into account that  the the trans-domain  (see Fig.
\ref{Schematic})  has the volume $V = (2\pi/3) R(t)^3$. In the dimensionless
notations
\begin{eqnarray}
 {\widetilde f}_{\rm count} (t) = \left[ \dfrac{3 C_2}{8(3\nu - 1)} \right]
\left( \dfrac{M (t)}{{\widetilde R}(t)^3}
\right)^{3\nu/(3\nu - 1)}.
\label{Osmotic}
\end{eqnarray}

The pore friction force (in the dimensionless units) ${\widetilde f}_{\rm pore}
(t) \equiv a f_{\rm pore}/T = 6 \pi \zeta_p a v (t)/T = (\zeta_p/\zeta_0)
{\widetilde
v}(t)$, where $\zeta_p$ is the pore friction coefficient and we have used the
notation ${\widetilde v}(t)  \equiv 6 \pi \zeta_0 a v(t)/T$. Taking into account
Eq.
(\ref{Velocity}) we have
\begin{eqnarray}
 {\widetilde f}_{\rm pore} (t) =  \dfrac{\zeta_p [{\widetilde
F}(t)]^{z-2}}{\zeta_0 {\widetilde X}(t)}.
\end{eqnarray}

Finally, by using Eq.~(\ref{F}) we obtain the algebraic equation
\begin{eqnarray}
{\widetilde F}(t) = {\widetilde f} -  \left[ \dfrac{3 C_2}{8(3\nu - 1)}
\right] \left[ \dfrac{M (t)}{{\widetilde R}(t)^3}
\right]^{3\nu/(3\nu - 1)} - \dfrac{\zeta_p [{\widetilde
F}(t)]^{z-2}}{\zeta_0 {\widetilde X}(t)}.
\label{Algebraic}
\end{eqnarray}

For  intermediate and strong forces (which corresponds to ``stem-flower''  and 
``stem'' scenario, respectively)  the pore friction reads ${\widetilde
f}_{\rm pore} (t) = (\zeta_p/\zeta_0) {\widetilde v}(t) =
(\zeta_p/\zeta_0){\widetilde F}(t) / {\widetilde X}(t) $, where we have used
Eq.~(\ref{Velocity_Stem_Flower}) (or analogously Eq.~(\ref{Balance_Force}) for the
``stem'' case). As a result, Eq.~(\ref{Algebraic}) will be replaced by 
\begin{eqnarray}
{\widetilde F}(t) = {\widetilde f} -  \left[ \dfrac{3 C_2}{8(3\nu - 1)}
\right] \left[ \dfrac{M (t)}{{\widetilde R}(t)^3}\right]^{3\nu/(3\nu - 1)} -
\dfrac{\zeta_p [{\widetilde
F}(t)]}{\zeta_0 {\widetilde X}(t)},
\label{Algebraic_Stem}
\end{eqnarray}
or, equivalently,
\begin{eqnarray}
 {\widetilde F}(t) = \dfrac{{\widetilde f} - \left[ \dfrac{3 C_2}{8(3\nu -
1)}\right] \left[ \dfrac{M (t)}{{\widetilde R}(t)^3}\right]^{3\nu/(3\nu - 1)}
}{1 + \dfrac{\zeta_p}{\zeta_0 {\widetilde X}(t)}}.
\label{Algebraic_Stem_1}
\end{eqnarray}

As a result we have four equations, i.e. Eqs. 
(\ref{Mat_Balance_Final}) , (\ref{Flux_of_Monomers}), (\ref{R}) and
(\ref{Algebraic}) (in the case of intermediate  forces these equations are Eqs.
(\ref{Material_Balance_Stem_Flower_Final}),
(\ref{Flux_of_Monomers_Stem_Flower}), (\ref{R})
and (\ref{Algebraic_Stem_1}))  for four unknowns ${\widetilde X}(t)$,
${\widetilde R}(t)$, $M(t)$ and ${\widetilde F}(t)$. The translocation 
problem within this model is treated {\it self-consistently}, which means 
that the resulting force in the pore $F(t)$ is not given but depends on the 
front positions on {\it cis}, ${\widetilde X}(t)$,  and  {\it trans}, 
${\widetilde R}(t)$, sides as well as on the number of  
translocated monomers $M(t)$. In the next section we
discuss the numerical solution of this set of equations. In doing so, we
will compare the results with the simplified case without crowding and pore
friction \cite{Dubbeldam}. A more detailed exposition for this case is given in Appendix
\ref{App_1}.

\section{Numerical computations}

The  resulting four equations, Eqs.~(\ref{Mat_Balance_Final}) ,
(\ref{Flux_of_Monomers}), (\ref{R}) and (\ref{Algebraic}), for the four variables: 
${\widetilde X}(t)$, ${\widetilde R}(t)$, $M(t)$ and ${\widetilde F}(t)$ are
known as the semi-explicit differential-algebraic equations (DAE)
\cite{Brenan}. For these particular DAE we can distinguish between the
differential variables: ${\widetilde R}(t)$ and $M (t)$, and the algebraic ones:
${\widetilde X}(t)$ and ${\widetilde F}(t)$. In the case of intermediate forces,
i.e.
$1 < {\widetilde f} < N^{\nu}$, the corresponding equations are Eqs. 
(\ref{Material_Balance_Stem_Flower_Final}),
(\ref{Flux_of_Monomers_Stem_Flower}), (\ref{R})
and (\ref{Algebraic_Stem_1}). By fixing the initial conditions for the differential
variables, ${\widetilde R}(0)$ and $M(0)$,  the corresponding initial
values for ${\widetilde X}(0)$ and ${\widetilde F}(0)$ are obtained from
Eqs.(\ref{Mat_Balance_Final}), (\ref{Algebraic}) employing the 
Newton-Ralphson method.  By alternatingly solving the differential equations for 
${\widetilde R}$ and $M$ (using the Euler forward method), and 
the algebraic equations for ${\widetilde X}(t)$ and ${\widetilde F}(t)$, 
we obtain the solution of the system of DAE. In all calculations the constant 
$C_2$, which naturally appears in the scaling expression for the free energy 
Eq.~(\ref{Delta_Free_Energy}),  has been  set to $C_2 = 1$. However, we verified that the 
numerical results do not change notably even when $C_2$ was set to $10$. 

\begin{figure}[ht]
 \includegraphics[width=8.0cm]{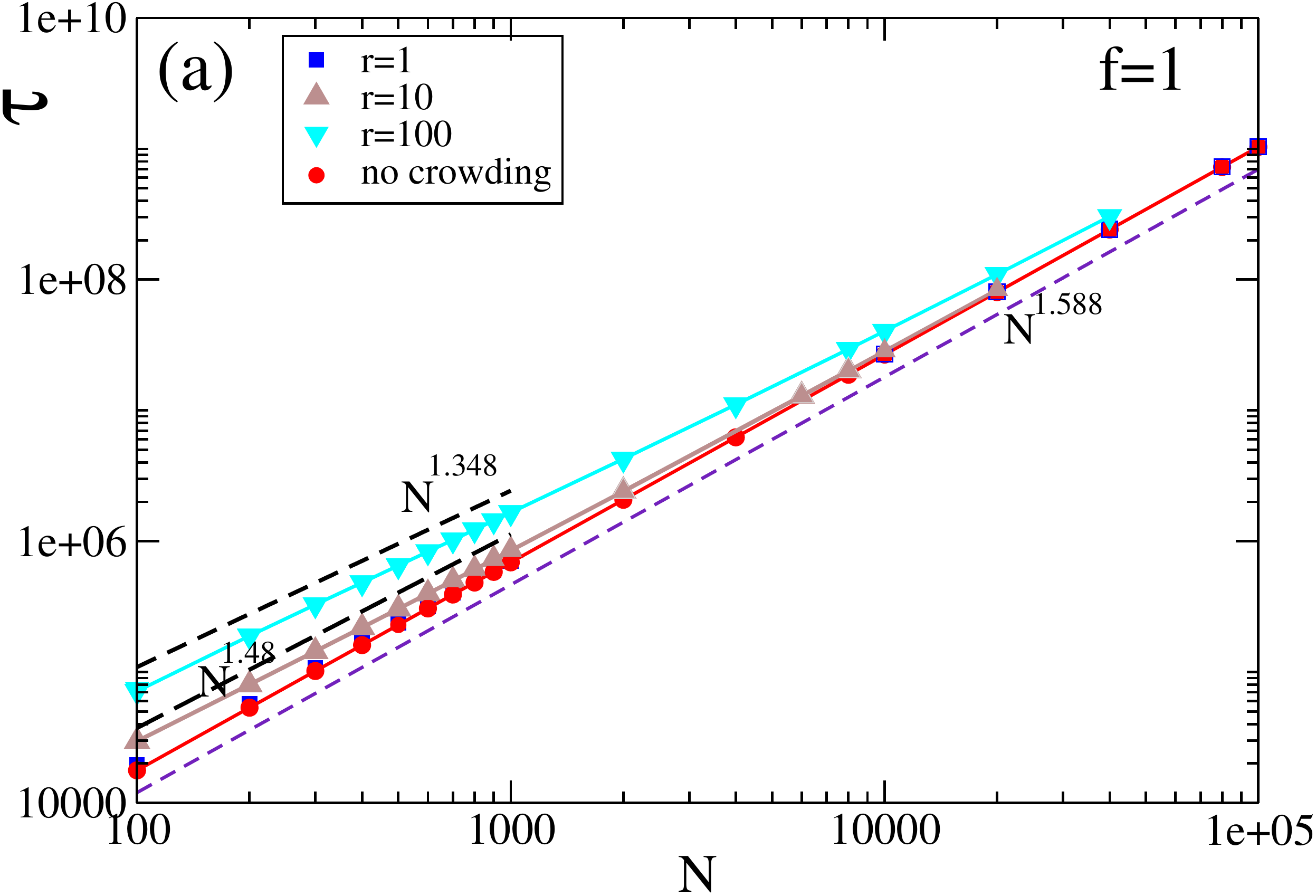}
\includegraphics[width=8.0cm]{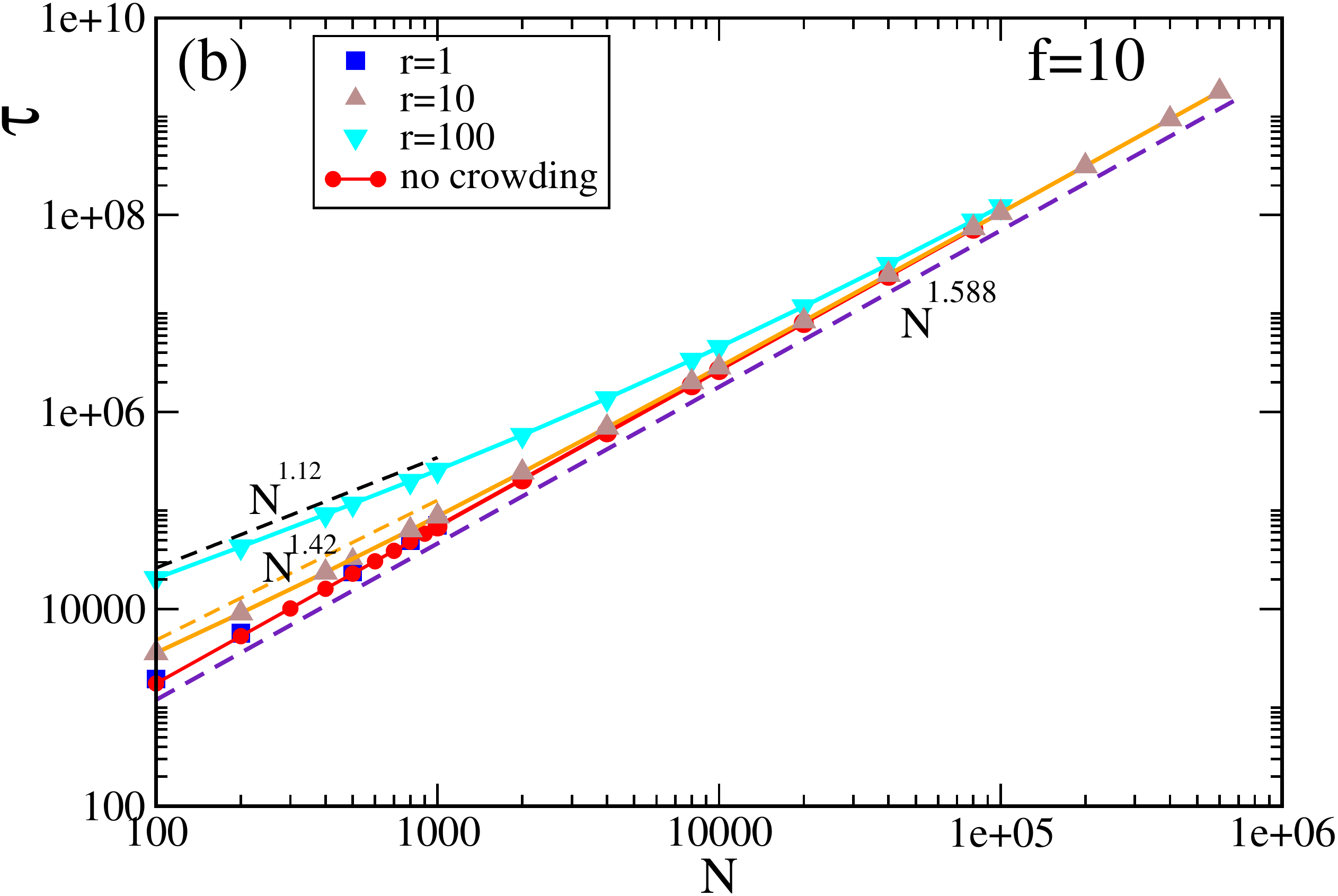}
\hspace{3cm}
\caption{Translocation time $\tau$ dependence on $N$ (with crowding and pore
friction effects) for different values of 
the forces: (a) ${\widetilde f} = 1$ and (b) ${\widetilde f} = 10$. The
different pore frictions ratios,  $r=\xi_p/\xi_0$, are specified in the legend. 
The scaling for the no crowding and no pore friction case  (outlined in 
Appendix \ref{App_1}) is shown for reference by filled  red round
symbols.}
\label{fig:fig2}
 \end{figure}

In Fig.~\ref{fig:fig2}, we show the translocation time $\tau$  vs.  chain 
length $N$ or two different driving forces $\widetilde{f}=1$ and $\widetilde{f}=10$.
We find scaling $\tau \propto N^{\alpha}$  in a wide range of chain lengths, $10^2 < N <
10^6$. We note that for ${\widetilde f} =1$ and ${\widetilde f}
=10$ we have used for calculations the ``trumpet''  and ``stem-flower'' 
scenarios,
respectively. The corresponding results are shown in Fig. \ref{fig:fig2}.  First
of all one can see that trans side crowding practically does not affect the 
scaling behavior: the curve corresponding to the case without crowding and pore 
friction (shown by red filled circles in Fig. ~\ref{fig:fig2}) coincides with 
the case with crowding but vanishing pore friction, i.e. $r = \xi_p/\xi_0  = 0$ 
(shown by blue boxes in Fig. ~\ref{fig:fig2}). In other words the impact of 
the ``crowding effect'' by itself  is almost negligible. Only for larger pore 
friction 
ratios, $r = 10$ and $r = 100$, the
exponent $\alpha$ mildly changes (especially for a relatively small force,
${\widetilde f} = 1$, shown in Fig. \ref{fig:fig2}a); larger values of $r$ correspond
to smaller values of $\alpha$. For the stronger force, ${\widetilde f} = 10$, the
value of the translocation exponent falls to $\alpha = 1.12$ for the high friction pore, 
$r=\xi_p/\xi_0 = 100$ ({\it cf.} Fig.\ref{fig:fig2}b ). This  value is smaller than
the  value of $\alpha$ for the no crowding case, $\alpha = 1 + \nu =1.588 $ (shown
by filled  red circles in Fig. \ref{fig:fig2}b ), and close to the linear 
scaling law, $\tau \propto N$, found experimentally by Kasianowicz {\it et al.}
\cite{Kasianowicz} for polyuridylic acid in the range of 100 - 500
nucleotides. On the other hand,  experiments  on double-stranded DNA 
translocation through a solid-state nanopore lead to the exponent $\alpha = 
1.27$ \cite{Storm,Storm_1}, which is close to our findings for ${\widetilde f} 
= 10$ and $10 < r < 100$ shown in Fig.\ref{fig:fig2}b.

Another interesting behavior exhibited by Fig. \ref{fig:fig2} is the finite
chain length effect due to the pore friction. One can see that as the chain
length $N$ increases the scaling exponent $\alpha$ approaches the ``no
crowding and no pore friction'' case, i.e. $\alpha = 1+\nu$ (see Appendix 
\ref{App_1}). Moreover, the larger the pore friction,
the greater is the chain length crossover $N_c$. For example, for ${\widetilde
f} = 10$ and $r = 100$ the crossover chain length $N_c \approx 10^5$. This
behavior is in full agreement with results of Ikonen {\it et al.}
\cite{Ikonen,Ikonen_2}.

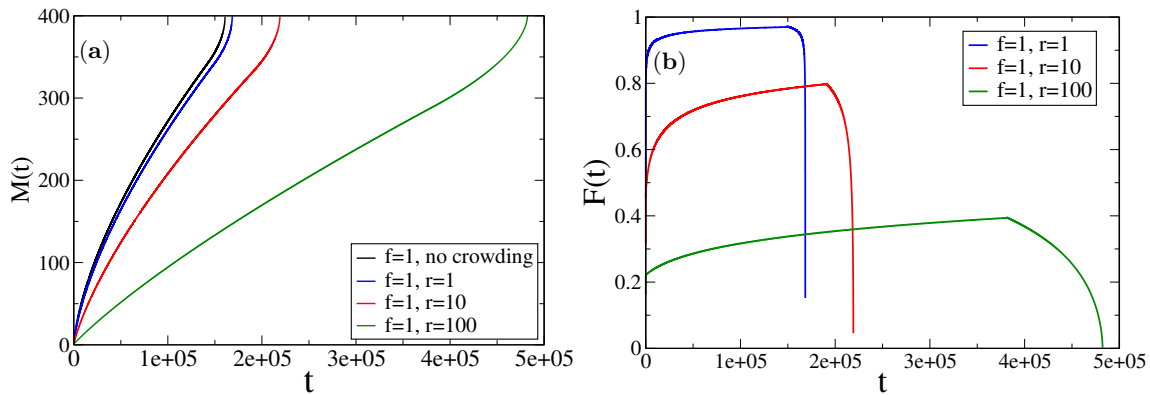
\begin{figure}[ht]
 \includegraphics[width=7.5cm]{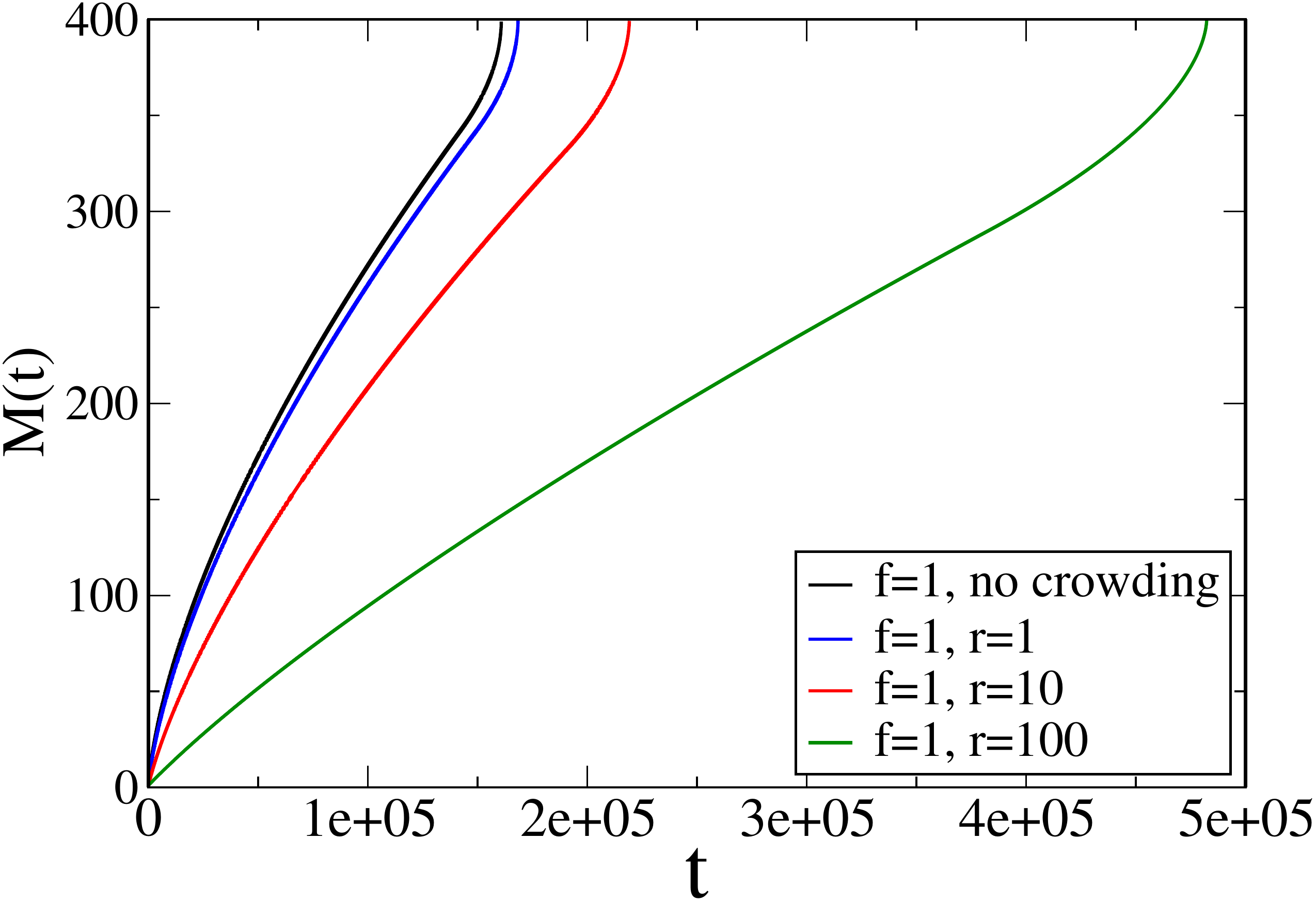}
  \includegraphics[width=7.5cm]{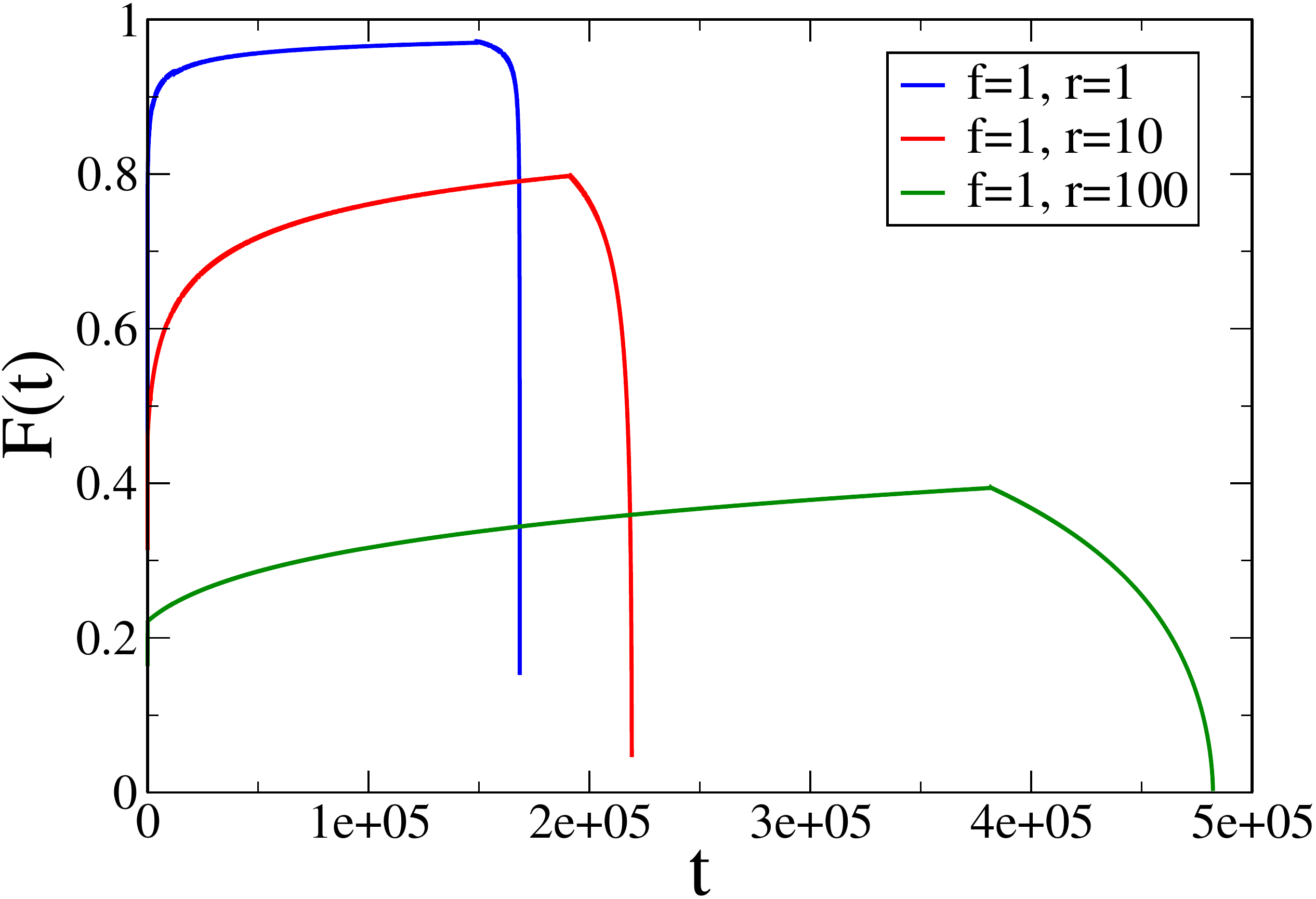}
\begin{picture}(0,0)
 \put(-407,127){$\bf (a)$}
 \put(-190,123){$\bf (b)$}
\end{picture}
\caption{In (a) the number of translocated monomers for a N=400 chain is shown
as  function of time. The upper curve corresponds to the sipmified ``no 
crowding and no pore friction case'' outlined in Appendix \ref{App_1}. (b) the 
dependence of the resulting  force 
in the pore as a function of time is displayed.}
\label{fig:fig4}
\end{figure}

Next we  investigate the dynamics of the translocation process. Figure
\ref{fig:fig4} shows the number of translocated monomers $M(t)$ and the
resulting force ${\widetilde F}(t) $ as functions of time. As one can see from 
Fig.~\ref{fig:fig4}a the
translocation velocity initially slightly decreases, however, when the chain has
nearly threaded the pore it experiences a  large acceleration as witnessed by 
the almost 
vertical tangent to the curve when $M(t)$ approaches the chain length $N=400$. 
In  Fig.~\ref{fig:fig4}a we compare in a clear way the simplified model without 
crowding and pore friction (as discussed in Appendix \ref{App_1}), on the one 
hand, and the model with crowding but without pore friction  friction ($r = 
0$), on the other hand. This comparison shows once more that  crowding by 
itself  hardly influences the speed of the translocation process.  
Alternatively, a large pore friction coefficient (as compared with the bulk
friction coefficient) leads to a clear dynamical slowing  down.  

The resulting force evolution given in Fig.~\ref{fig:fig4}b  first shows a
gradual increase which after reaching its maximum value rapidly decreases to
zero. It is interesting that the maximum of force is attained when the
propagating front on the {\it cis} - side has reached the end of the 
polymer chain (tension propagation stage).  After that the whole chain, which
participates in the translocation process, starts to accelerate.
During this stage (known as the tail retraction stage \cite{Ikonen_1,Ikonen_2})
the tensile force in the chain starts to drop and vanishes when the chain has
fully translocated through the pore.

\vskip 1.0cm

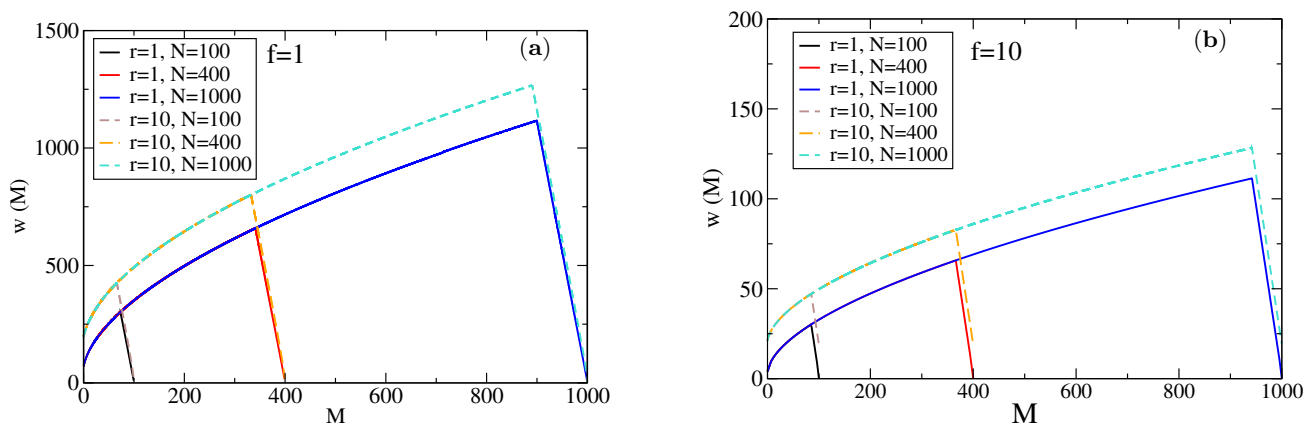
\begin{figure}[ht]
 \includegraphics[width=8.0cm]{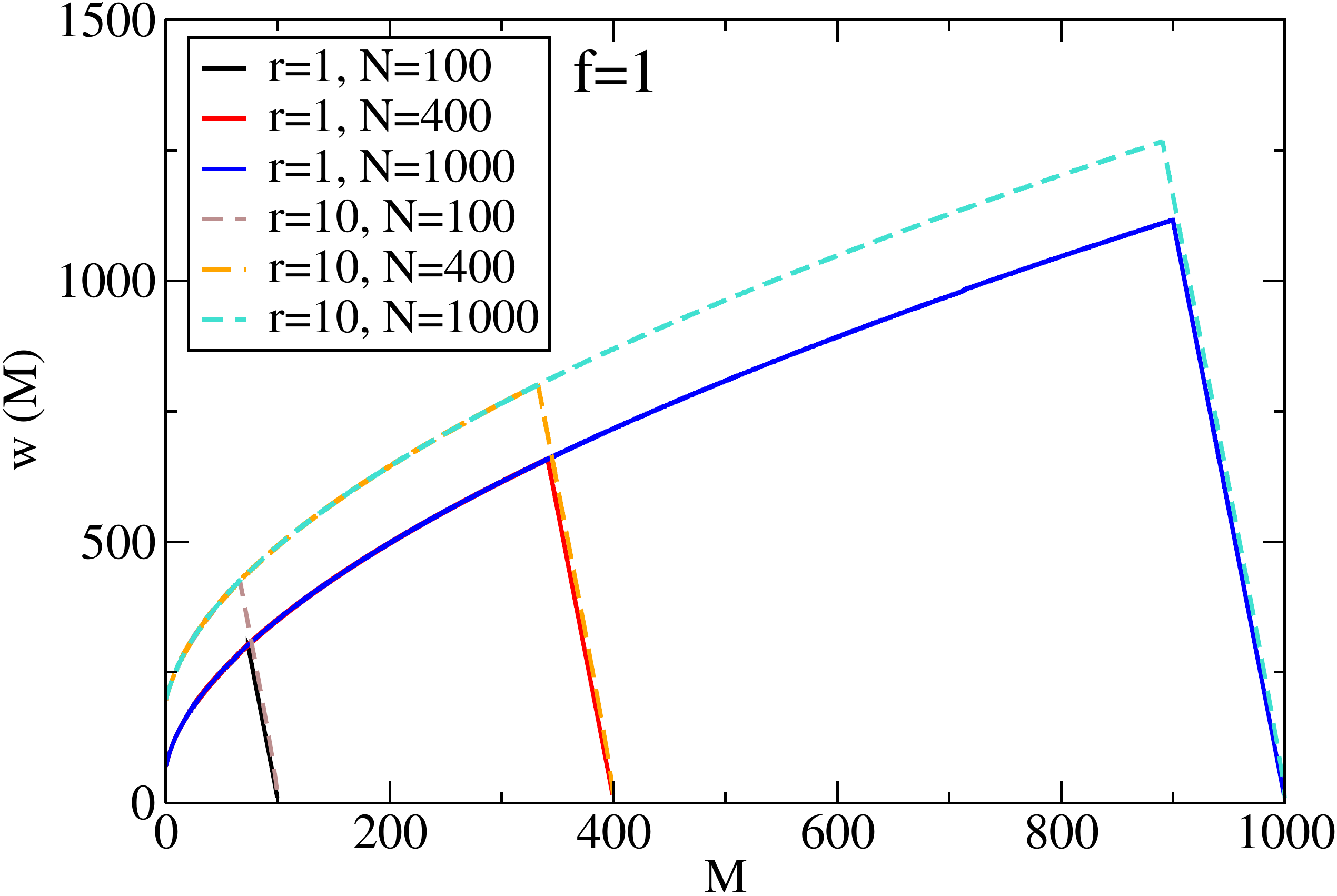}
\hspace{1cm}
\includegraphics[width=8.0cm]{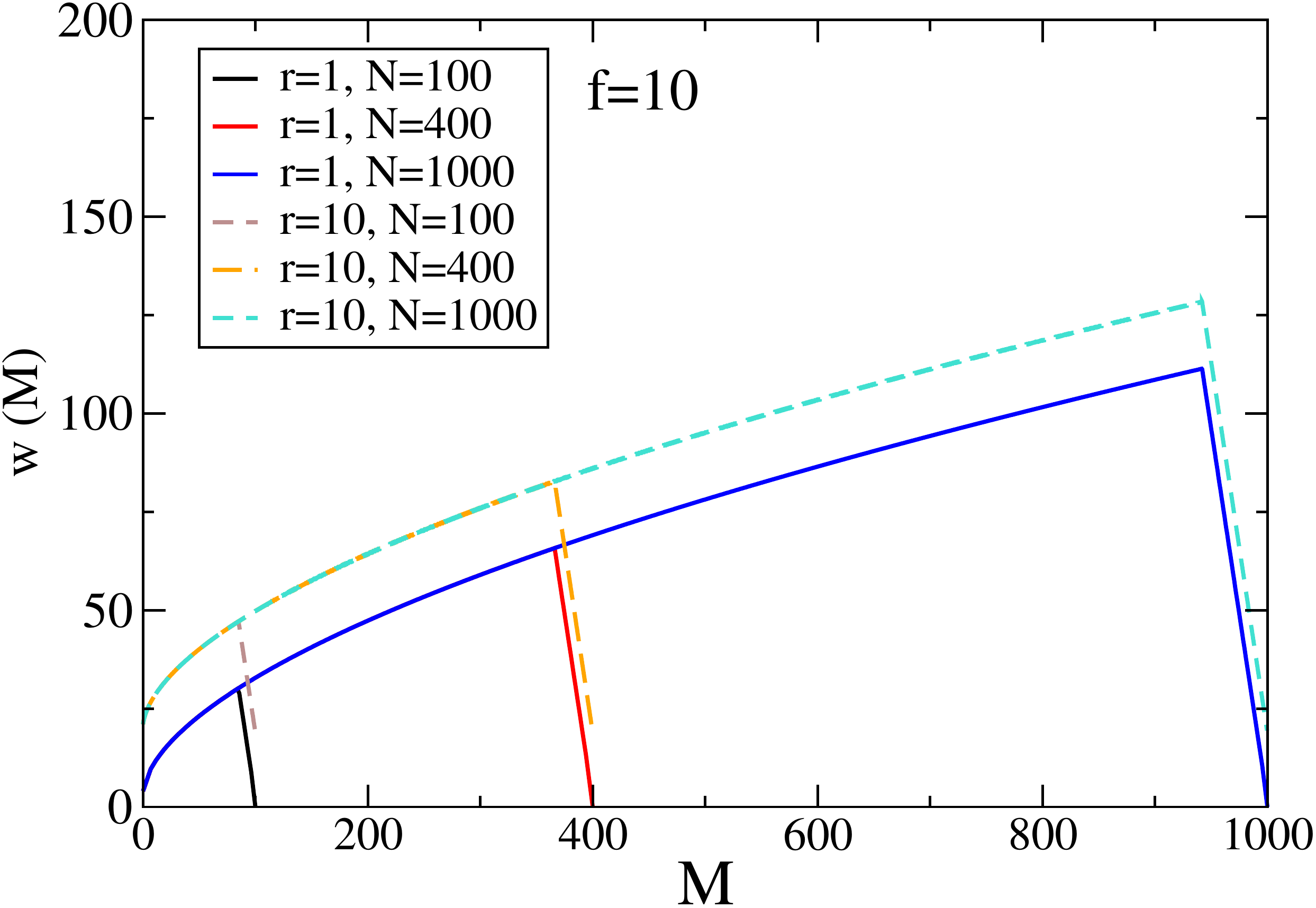}
\begin{picture}(0,0)
 \put(-300,140){$\bf (a)$}
 \put(-45,143){$\bf (b)$}
\end{picture}
\caption{Waiting time distribution function, $w(M) = (d M/d t)^{-1}$, as a
function of  $M$ for different forces: a) $f=1$ and b)~$f=10$. Chain lengths,
$N=100, 400, 1000$, and friction coefficients ratios, $r = \xi_p/\xi_0 = 1, 10$,
are shown in legends.}
\label{fig:waiting_time}
 \end{figure}

The same two stages of translocation also could be seen on the waiting time
distribution $w(M)$ which is defined as the time that takes for the transition
$M \rightarrow M + \Delta M$ ({\it cf.} ref. \cite{Linna_3,Ikonen_1,Ikonen_2}).
It is apparent that in the continuous limit the waiting time distribution is
nothing but the inverse translocation coordinate velocity, i.e. $w(M) = (d M/d
t)^{-1}$. In Figure \ref{fig:waiting_time} we show these distributions for  different chain lengths, pore frictions and forces. 
Again one can discern the tensile force propagation stage during which 
transloaction slows down, which is then followed by the chain tail retraction stage during  
which the translocation process speeds up. Our results are in qualitative agreemenent with the
findings based on MD-simulation and BDTP-model
\cite{Linna_3,Ikonen_1,Ikonen_2}).

\section{Conclusions}

We have given a detailed theoretical interpretation of crowding  and pore
friction effects  in the course of driven polymer
translocation. Translocation dynamics is treated  
self-consistently   when the resulting force in the pore, $F(t)$,  
depends on the front positions on {\it cis}, $\widetilde X(t)$, and {\it trans}, 
$\widetilde R(t)$, sides as well as on the number of translocated monomers 
$M(t)$.  This approach provides a generalization of the BDTP-model 
\cite{Ikonen,Ikonen_1,Ikonen_2} where  the time-dependent friction  coefficient 
of the {\it cis} side moving domain  was taken from the simplified model 
without crowding  and pore friction.
The resulting four  differential - algebraic equations for four dynamical 
variables, 
${\widetilde X}(t)$, ${\widetilde R}(t)$, $M(t)$ and ${\widetilde F}(t)$, were
derived by taking into account the tensile force propagation on the {\it cis} -
 side of the membrane as well as the concentration blob picture on the {\it trans}-side. 
Our detailed numerical solutions of these equations show that the
translocation dynamics is scarcely affected by the  crowding itself, which is
consistent with previous findings \cite{Ikonen_1,Ikonen_2}. On the other hand,
in the presence of  pore friction the translocation process not only becomes
slower but also the translocation exponent $\alpha$ (especially for relatively
large driving forces) decreases as compared to the idealized case without
crowding and pore friction, i.e. $\alpha < 1 + \nu$. With increase of
 chain length the translocation scaling  asymptotically approaches
the ``no crowding no pore friction'' case, i.e. the scaling exponent $\alpha = 
1 + \nu$ (see Appendix~\ref{App_1}). The
 crossover is very broad with the corresponding critical chain length $N_c
\approx 10^5$ (for large force and high pore friction). This conclusion is in a
full agreement with results of Ikonen {\it et al.} \cite{Ikonen,Ikonen_2}. 
Hence  the translocation exponent $\alpha$ is not  universal  and (for 
relatively short polymer 
chains and strong forces)  mainly  the \emph{pore friction} 
could lower its value. This, in turn, explains 
the large variety of  $\alpha$
values which have been reported in experiments  \cite{Kasianowicz,Storm,Storm_1}
and computer simulations \cite{Milchev}.

\section*{Acknowledgement}
 We would like to thank A.Y.~Grosberg, R.P.~Linna and  A.~Milchev for fruitful
discussions.
 V.G.~Rostiashvili acknowledges support from the Deutsche
Forschungsgemeinschaft (DFG), grant No. SFB 625/B4.

\begin{appendix}
\section{Simplification: no crowding and no  pore friction}
\label{App_1}

Let's now simplify matters by neglecting crowding and pore friction effects. In
this case the effective force ${\widetilde F} (t) = {\widetilde f}$ and we go
back to the case which was investigated in ref. \cite{Dubbeldam}. Then 
Eqs.~(\ref{Flux_of_Monomers}) and 
(\ref{Mat_Balance_Final}) become closed and we have
\begin{eqnarray}
\dfrac{d M(t)}{d{ \widetilde t}} = \dfrac{{\widetilde f}^{z-1 -
1/\nu}}{ 6\pi \: {\widetilde X}(t)} 
\label{M_Simplified}
\end{eqnarray}
and
\begin{eqnarray}
 \dfrac{{\widetilde X}(t)}{ C_1 \: {\widetilde f}^{\frac{1}{\nu} - 1}} +
M(t) = [{\widetilde
X}(t)]^{\frac{1}{\nu}}
\label{Mat_Balance_Simplified}
\end{eqnarray}
At $t > \tau_1$ instead of the material balance equation  Eq.
(\ref{Mat_Balance_Simplified}) we have
\begin{eqnarray}
 \dfrac{{\widetilde X}(t)}{ C_1 \: {\widetilde f}^{\frac{1}{\nu} - 1}} +
M(t) = N
\label{Mat_Balance_1_Simplified}
\end{eqnarray}
This simplified case enables to solve the problem analytically. Really, after
differentiation of Eq. (\ref{Mat_Balance_Simplified}) and combination with Eq.
(\ref{M_Simplified}) we have 
\begin{eqnarray}
 \dfrac{1}{\nu} \left[  1 - ({\widetilde f} {\widetilde X})^{1/\nu - 1}
\right]\: \dfrac{d{\widetilde X}}{d \: {\widetilde t}} = - \dfrac{{\widetilde
f}^{z-2}}{6\pi {\widetilde X}}
\end{eqnarray}
where we have also used that for the Rouse model $C_1 = \nu$. This equation
could be easily solved with the natural initial condition ${\widetilde X}(0) =
1/{\widetilde f}$. The corresponding solution reads
\begin{eqnarray}
 {\widetilde t} = t_0 + \dfrac{6\pi}{1 + \nu} \; {\widetilde f}^{1/\nu - z + 1}
{\widetilde X}^{1/\nu + 1} \left[ 1 - \dfrac{1 + \nu}{\nu ({\widetilde
f} {\widetilde X})^{1/\nu - 1}}\right]
\label{X_vs_time}
\end{eqnarray}
where $\tau_0 = 6\pi/ [\nu (1 + \nu) {\widetilde f}^z]$.
The first stage of the translocation process, tension propagation, is continued 
up to $t = \tau_1$ when the tension attain the very last monomer, i.e.
${\widetilde X} (\tau_1) = N^{\nu}$. In this case the characteristic
(dimensionless) time of the first stage reads
\begin{eqnarray}
\tau_1 = \dfrac{6\pi }{1 + \nu} \; {\widetilde f}^{1/\nu - z + 1}
N^{1 + \nu} \left[ 1 - \dfrac{1 + \nu}{\nu ({\widetilde
f} N^{\nu})^{1/\nu - 1}}\right]
\end{eqnarray}

At the second stage, tail retraction, the material balance equation is given
by Eq. (\ref{Mat_Balance_1_Simplified}), which, due to Eq.
(\ref{M_Simplified}), yields
\begin{eqnarray}
 {\widetilde X} \: \dfrac{d {\widetilde X}}{ d {\widetilde t}} = -
\dfrac{\nu }{6\pi}\:  {\widetilde f}^{z - 2}
\label{Second_Stage_X}
\end{eqnarray}
As it can be seen at the tail  retraction regime  $d {\widetilde X}/ d
{\widetilde t} < 0$ and ${\widetilde X}$ decreases from ${\widetilde X}(\tau_1)
= N^{\nu}$ up to ${\widetilde X}(\tau_{\rm fin})= 0$, where $\tau_{\rm fin}$ is
the final time moment of the translocation. The solution of Eq.
(\ref{Second_Stage_X}) has the form
\begin{eqnarray}
 {\widetilde t} = \tau_{\rm fin} - \dfrac{3 \pi {\widetilde
X}^2}{\nu {\widetilde f}^{z - 2}}
\label{Second_Stage_Solution}
\end{eqnarray}
The second stage of translocation lasts $\tau_2 = \tau_{\rm fin} - \tau_1$ and
due to Eq. (\ref{Second_Stage_Solution}) we have
\begin{eqnarray}
 \tau_2 \propto \dfrac{N^{2\nu}}{{\widetilde f}^{z - 2}}
\end{eqnarray}
As a result the total traslocation time is given as \cite{Dubbeldam} 
\begin{eqnarray}
 \langle \tau \rangle &=& \tau_1 + \tau_2 \nonumber\\
&=& B_1 \dfrac{N^{1+\nu}}{{\widetilde f}^{z - 1 - 1/\nu}} + B_2
\dfrac{N^{2\nu}}{{\widetilde f}^{z - 2}}
\end{eqnarray}
For the Rouse case (i.e. when $z = 2 + 1/\nu$) we arrive at the result
 \begin{eqnarray}
  \langle \tau \rangle = B_1 \dfrac{N^{1+\nu}}{{\widetilde f}} + B_2
\dfrac{N^{2\nu}}{{\widetilde f}^{1/\nu}}
\label{eq:rouseres}
 \end{eqnarray}
This important result, which indicates the crossover between the tension 
propagation and tail retraction regimes as the chain length $N$ increases,  
has been derived for the first time in ref.
\cite{Dubbeldam} (see Eq. (2.26) in this reference) by a slightly different
method. Eq. (\ref{eq:rouseres}) predicts that the effective translocation
exponent falls in the range $2\nu \leqslant \alpha \leqslant 1+\nu$, which
closely agrees with MD-findings by Luo {\it et al.} \cite{Luo}. This crossover 
could also lead, along with polymer-pore friction, to  lower values of the 
translocation exponent $\alpha$ for relatively small chain lengths.

Lastly, we should mention that the result given by Eq. (\ref{eq:rouseres}) 
leads to the correct scaling for the unbiased translocation case when the 
driving force ${\widetilde f}$ is vanishingly small. Really, in this case 
${\widetilde f} \longrightarrow 1/N^{\nu}$ (this is the lower bound for the 
Pincus blob formation \cite{DeGennes}) and Eq. (\ref{eq:rouseres}) can be 
written as $ \langle \tau \rangle  \propto N^{2\nu +1}$. This result was 
obtained first by Kantor\&Kardar \cite{Kantor} in a different way.

\end{appendix}

\end{document}